\documentstyle[aps,prl,twocolumn,graphicx,floats]{revtex}
\begin{document}
\draft
\wideabs{
\title{Formation of helical states in wormlike polymer chains}
\author{Josh P. Kemp and Zheng Yu Chen}
\address{ Guelph-Waterloo Program for Graduate Work in Physics
Department of Physics, University of Waterloo
Waterloo, Ontario, Canada N2L 3G1}
\date{\today}
\maketitle
\begin{abstract}
	We  propose a potential for wormlike 
polymer chains which can be used
to model the low-temperature conformational structures. 
We 
successfully reproduced helix ground states up to 6.5 helical loops,
 using the multicanonical Monte Carlo simulation method.
We demonstrate that
the coil-helix transition involves four distinct phases: 
coil(gaslike), 
collapsed
globular(liquidlike), 
and two helical phases I and II (both solidlike). 
The helix I phase is characterized by a helical 
structure with dangling loose ends, and the
helix II phase corresponds to a near perfect helix ordering in 
the entire crystallized chain.
 

\end{abstract}
}

The structure of $\alpha$-helices is one of the most 
common conformations occurring in
biological molecules such as proteins. 
It is this unique structure that acts as
the building block for much more 
complicated tertiary structures \cite{Biobook}. 
In 1959, Zimm and Bragg proposed a mean-field model for 
the  coil-helix transition that 
was based on  
the thermodynamics of
a known sequence of pairing residues\cite{ZB}. 
A tremendous number of
theoretical works have been published since then 
and the basic approaches are derivatives or 
alterations of the original Zimm-Bragg treatment \cite{Theories,Qian}.
%
Most recent numerical attempts to study
the problem are based on more realistic  potentials at the atomic level
for helix-forming polypeptides,   concentrating more on finding
the ground state structure \cite{Sims}, than on
the thermodynamics at finite temperatures (see, however, Ref. \cite{Oka}). 

From the statistical-physics perspective, it is desirable to start from 
a simple theoretical model of general characteristics,
in the same sprite as
recent models for  double stranded
semiflexible polymers \cite{Liv}, for low-temperature lattice homopolymers \cite{Zhou} 
and for polymer chains containing dipolar segments \cite{Muthu}.
%
%
By ignoring the specifics of the amino acids,
most proteins can be viewed as wormlike polymer chains.
One can then ask simple questions such as, what are the minimum conditions
required for a wormlike polymer chain to reproduce
a helix ground state? What are the characteristics
of the so-called coil-helix transition in comparison with the predictions of a 
 Zimm-Bragg-type theory, and
how different is it from the coil-collapsing 
transition in other polymeric molecules
occurring near the theta temperature?
From a polymer physics point of view,
a related and inspiring question by itself is:
what are the low temperature structures of a wormlike polymer chain
in general?

There are two basic characteristics common to all wormlike chains: 
the existence of a finite persistence length,
which prevents the polymer from folding as a sharp 
hairpin loop at scales smaller than the persistence length, and   
an excluded volume
interaction between 
the monomers, which prevents polymer chains from collapsing into
a point at low temperatures.
For helix-forming wormlike polymers, 
a third basic characteristic must be included: 
a
directionally biased attractive potential.


In the current approach we model  wormlike 
polymers by fixing the bond angle
$\theta$ between  two adjacent polymer bonds of length $a$, 
and allowing the azimuthal angle $\phi$ rotate freely,
as prescribed in the original freely rotating
 model \cite{Yamakawa}.
Movement of monomers is 
realized by implementing a  pivot
algorithm \cite{Pivot}, where 
a bond is chosen at random as a reference axis about which the 
entire polymer segment attached to this bond is rotated.
The Metropolis method of sampling a canonical phase space
is known to be ineffective
for 
examining the possible conformational states 
at low temperatures, as a particular configuration 
could easily become trapped 
in a local energy minimum. 
To avoid this problem 
we have incorporated a multicanonical technique into the pivot 
algorithm, which involves reweighting the 
temperature as a function of energy, in order to
produce a flat simulation 
histogram across the entire energy landscape \cite{Sims,Uli}.
This approach
allows the polymer to potentially
visit all possible energy states and to tunnel through  
local energy minima. 
For a more detailed explanation of the
multicanonical technique 
the reader is referred to Refs \cite{Oka,Uli}. 

The interaction energy
between the monomers labeled $i$ and $j$ having a center-to-center distance $r_{ij}$ 
is proposed to have the simple form,
\begin{equation}
{V}_{ij}=\cases {
0
&
$ {\rm for~} \sigma \le r_{ij}   $
\cr
{-\epsilon [{\vec u}_i\cdot ({\vec r}_i -{\vec r}_j)]^m -\epsilon [{\vec u}_j\cdot ({\vec r}_i -{\vec r}_j)]^m} 
&  
$ {\rm for~} d \le r_{ij} \le \sigma   $\cr
\infty
&
$ {\rm for~} 0 \le r_{ij} \le d $\cr
}
\end{equation}
where $\vec u_{\rm i}$ is the cross product defined as
$$
\vec u_{\rm i} = (\vec r_{\rm i} - \vec r_{\rm i-1})\times (\vec r_{\rm i+1} - \vec
r_{\rm i}).  \eqno   (2)
$$
Here,
 $m$ represents the degree of directional bias
in the potential, 
$d$ is the excluded-volume diameter, and $\sigma$ is 
the  attractive-force range.
The multicanonical
technique is also an effective method to search for the
ground states of a given potential.
In an effort to answer the question of 
what type of potential  is needed for the creation of a perfect
helical ground state, the consequences of using various
values of $m$ were examined. 
We found that for an isotropic interaction ($m=0$), the stable 
ground state structure is a disordered globular state
with no particular relative directional ordering between the 
chain bonds. Thus,
a directionalized interaction ($m\neq 0$) is obviously an 
essential condition 
for a helix-forming wormlike chain.
We also found that weaker powers ($m=2,4$), 
would only produce a perfect helix ground state
in short polymers; in longer polymers 
  multiple bonds formed by spooling around a
helix core i
might give rise to an even lower 
energy.
In a helix state,
bonding effects should be maximized when all monomers
siting directly on top of each other,
demanding a strongly directionalized
interaction potential as in 
known helix-forming polypeptides.
%
As an illustration of some of the 
specific physical properties of a helix-forming wormlike chain,
we  report   here the case of $m=6$ only. Numerical data collected
on using other power $m$
will be reported  elsewhere.

%
It is worth noting that
our potential is not simply a function of the distance between the two involved monomers.  
The bonding preference also involves the directions of
the bonds  connected to these two monomers,   
which is 
an interesting effect
in the transition to the helix states, similar to
 a protypical hydrophobic interaction  in proteins. 
There are a number of other variables in the potential model.
The diameter $d$ and constant $\sigma$ 
have been adjusted to have the values  $3/2a$ and ${\sqrt {45/8}}a$,
while the  nearest neighbor  interaction  is ignored.
The fixed bond angle 
$\theta$ is chosen to have a value of approximately $\pi/3$, 
corrected for the helical pitch to  maintain six monomers per helix loop.
These selections of values
were based on the observation 
that under the current choice a stray end of the
monomers cannot loop through an open core 
of a partially formed helical segment. 
We have also investigated the case
of $\theta=\pi/5$, and found no major qualitative differenced in
the physical properties to be examined below.
This choice of 
potential, with the use multicanonical technique, was
able to produce a perfect helical ground state relatively quickly 
(see Fig 1d) for various polymer lengths $N=13,19,26$
and $39$, showing
that the helix ground state can be constructed with the above mentioned
three conditions of a helix-forming wormlike chain. 
\begin{figure}
\begin{center}
\includegraphics[height=8cm,width=7.5cm]{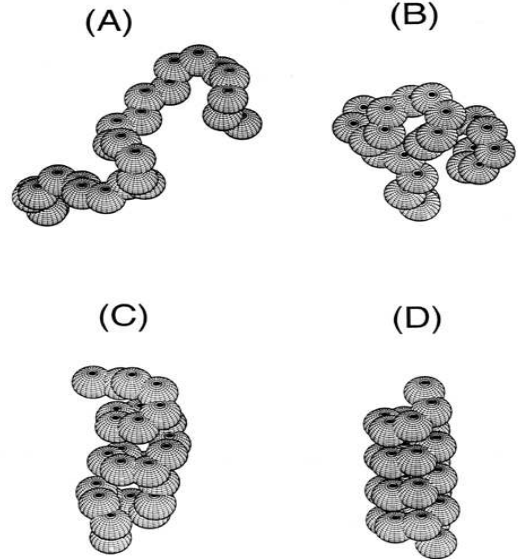}
\caption[width=5cm]{Snapshots of the configurations of a 26mer
at various temperatures: (a) $k_BT/\epsilon = 4$ (coil), 
(b) $k_BT/\epsilon = 1.3$  (globular) (c) $k_BT/\epsilon = 0.8$ (helix I)
  and
(d) $k_BT/\epsilon = 0.1$ (helix II).  The size of the beads represents the
actual hard-core interaction between non-adjacent monomers, and
 the attraction force range is ${\sqrt {5/2}}a$,
where $a$  is the bond length. A highly directionalized potential
with $m=6$ (see Eq. (1)) is used.}
\label{Pic1}
\end{center}
\end{figure}

Therefore, a wormlike chain can have numerous types of 
ground states,
which are dependent on the type of directional potential used. 
The construction a potential which produces 
a singular ground state of a desired
conformation is possible, although not a trivial exercise. 
The use of multicanonical
technique makes the search for these specific ground states simpler. 
Knowing that our model has an attainable perfect helix
ground state, we have calculated the multicanonical reweighting 
function 
for the 
temperature range discussed below, which enable us to 
conduct measurements of physical quantities in the
subsequent production runs. As a most effective 
measure  for 
exploring the phase transition signatures, 
the specific heat capacity of the polymer chain
was calculated by measuring the energy fluctuations 
$$ 
C_v/k_B = {{\langle E^2\rangle -\langle E\rangle ^2}\over{ (k_BT)^2}} 
\eqno (3)$$
The mean square radius of gyration, 
$$ 
<R_{\rm g}^2>=<{1\over N}\sum_{i=1}^N (\vec r_{\rm i} -\vec r_{\rm cm})^2 >
\eqno (4)$$ 
where $\vec r_{\rm cm}$ is the center of mass vector, and 
 the mean square end to end distance,  
$$ 
<R_{\rm end}^2>= <(\vec r_{\rm 1} -\vec r_{\rm N})^2 > \eqno (5)
$$ 
were also calculated to characterize the spatial dimension 
the polymer chain.
Finally we introduced three parameters to measure the degree of helicity
within the chain.
This is accessed in many references 
by examining the Ramachandran 
angles $\phi$ and $\psi$ which are parameters not
present in our current
model. Equivalent to the same treatment, 
the first helicity parameter, $H_1$, 
is based on an examination of 
the distance between 
the $i^{th}$ and $(i+3)^{th}$ monomers, which
is calculated and
evaluated as being in a helical or non-helical
 state, as the three connected bonds in the segment
will be arranged in a very specific way when in the helix state.
A window of allowed
distances can then define a three-monomer
segment as helical, and the average number of helical
segments, $H_{\rm 1}$, can be measured. A
second method of characterizing the degree of local helicity
within the polymer involves summing 
the dot products of the adjoining 
cross products of connected bond 
vectors: 
$$ 
H_{\rm 2} = {1\over N-2} \sum_{i=2}^{N-1} (\vec u_{\rm i} \cdot \vec u_{\rm i+1}) 
\eqno (6)$$
A third measurement is 
the correlation function
$G(i) \equiv {\vec u_i}  \cdot {\vec  u}_{\rm mid}$,  
where $\vec u_{\rm mid}$ corresponds to the vector in Eq. (2) for
the central monomer,
characterizing the correlation of the helical ordering along the
chain. Discussed 
below is the integrated correlation function,
$$ 
H_{\rm 3} = {1\over N-2} \sum_{i=2}^{N-1} 
(\vec u_{\rm i} \cdot \vec u_{\rm mid}) \eqno (7)
$$ 
which represents the averaged correlation in the chain. 
These three 
parameters yield a global representation of the
amount of helicity in the polymer.

Using the reweighting function from   preliminary
deterministic multicanonical runs, we performed
a typical production
run of $5 \times 10^8$ pivot rotations 
with measurement made every 10 steps  for rescaled 
temperatures $k_BT/\epsilon$
ranging from 0.1 to 5.0. 
Polymer chain lengths of
13,19,26, and 39 monomers were considered 
and the results are shown in Fig. 2.

Displayed in Fig. 2a, the scaled heat capacity curves 
shows 
a number of interesting
features.
 The first and most obvious is the large peak near
$kT/\epsilon=1.2$ corresponding to the
transition to a helix state.
The words describing what is meant by a helix state must be chosen
carefully for the reason that there are two distinct 
helix states observed in this study.
Figures 1b and 1c are  typical configurations just before
and after the transition.
Corresponding to a significant increase in the three helical
parameters measured for $N=19$ (Fig. 2c), a
 $C_v$ anomaly occurs
at this temperature, 
confirming that the polymer is making a
transition from a disordered (Fig. 1b) state  
to the more ordered helical state
(Fig. 1c). The same sharp increase in $C_v$
was also predicted in the Zimm-Bragg theory \cite{ZB,Biobook} 
and its variations, and in recent MC simulations by Okamoto and Hansmann \cite{Oka}. 
Zimm and Bragg argued
that this peak  does not represent a second order transition
in consistent with the known conclusion regarding the 
impossibility of phase transitions in a one dimensional system. However,
one can also argue that a helix polymer
is not strictly a one dimensional system.
Our peaks in heat capacity curves
continuously increase as $N$ grow, which show a similar behavior as
heat capacity curves in a protypical second order phase transition, 
and 
qualitatively agree  with the observation in a more realistic polypeptide model for 
shorter helical loops \cite{Oka}.
It would be interesting to further 
examine a possible finite size scaling
in the $C_v$  peak values determined from our
simulations. However, currently, the main obstacle to  achieving this 
goal is the increasing computational time for larger
$N$.

One of the most interesting features of Fig. 2a is 
the previously unknown
second transition occurring in the neighborhood of scaled
temperature $k_BT/\epsilon=0.3$. 
In a recent study of the phase transitions in
 isotropic homopolymers with specific ground states 
in a lattice model, 
Zhou et al. \cite{Zhou}  have observed a similar transition,
and have suggested 
that this is a solid-solid transition accompanying
the crystallization of their polymers into the ground state. 
In the present
case, our wormlike chains have the specifically
constructed ground state of a perfect 
helix, and we believe that the second transition corresponds
 to the crystallization 
of the helix I state into a perfect helix II state.
Above this transition we have observed that 
the chain is in a 
highly ordered helical structure only segmentally as demonstrated by Fig. 1c.
In particular,
the end portion of the polymer may not
necessarily ordered. 
Even for the ordered segments, elongation motion and bending of the ordered segment
 as a whole would make the directional ordering weaker
 compared with the more tightly wound helix II. 
Thus, 
as the temperature continues to drop further, 
the few unbonded end
segments arrange themselves into a perfect helical state (helix II),
and in the mean time, the already existing helix segments
crystallize to a near ground-state structure, 
 shown in Fig. 1d. Such a crystallization transition has not 
been  seriously modeled  by previous theoretical studies, 
which usually treat the helix I to 
helix II transition as a smooth crossover with no 
anomalies in the heat capacity, as  a result of 
influence by Zimm-Bragg's original model \cite{Biobook,ZB,Theories,Qian,Note}.
%
The three lower curves in Fig. 2c are 
derivatives of the helical parameters,
all 
showing  a significant
change in the slope corresponding to the low-temperature peak 
at the $C_v$ curve.
%
To further confirm this crystalline transition
we have examined individual local 
 the helical parameters
${\vec u}_i  \cdot {\vec u}_{i+1}$, 
which also show substantial changes
before and after the helix I  to helix II transition.
As  $N$ becomes large, the peak  position in $C_v$  
seems to be to move to higher temperatures, 
with the $C_v$ curves becoming flatter.
Presently, we  are unable to confirm these trends due to 
the large error bars associated with the low temperature behavior.
\begin{figure}
\begin{center}
\includegraphics[height=17cm,width=8.5cm]{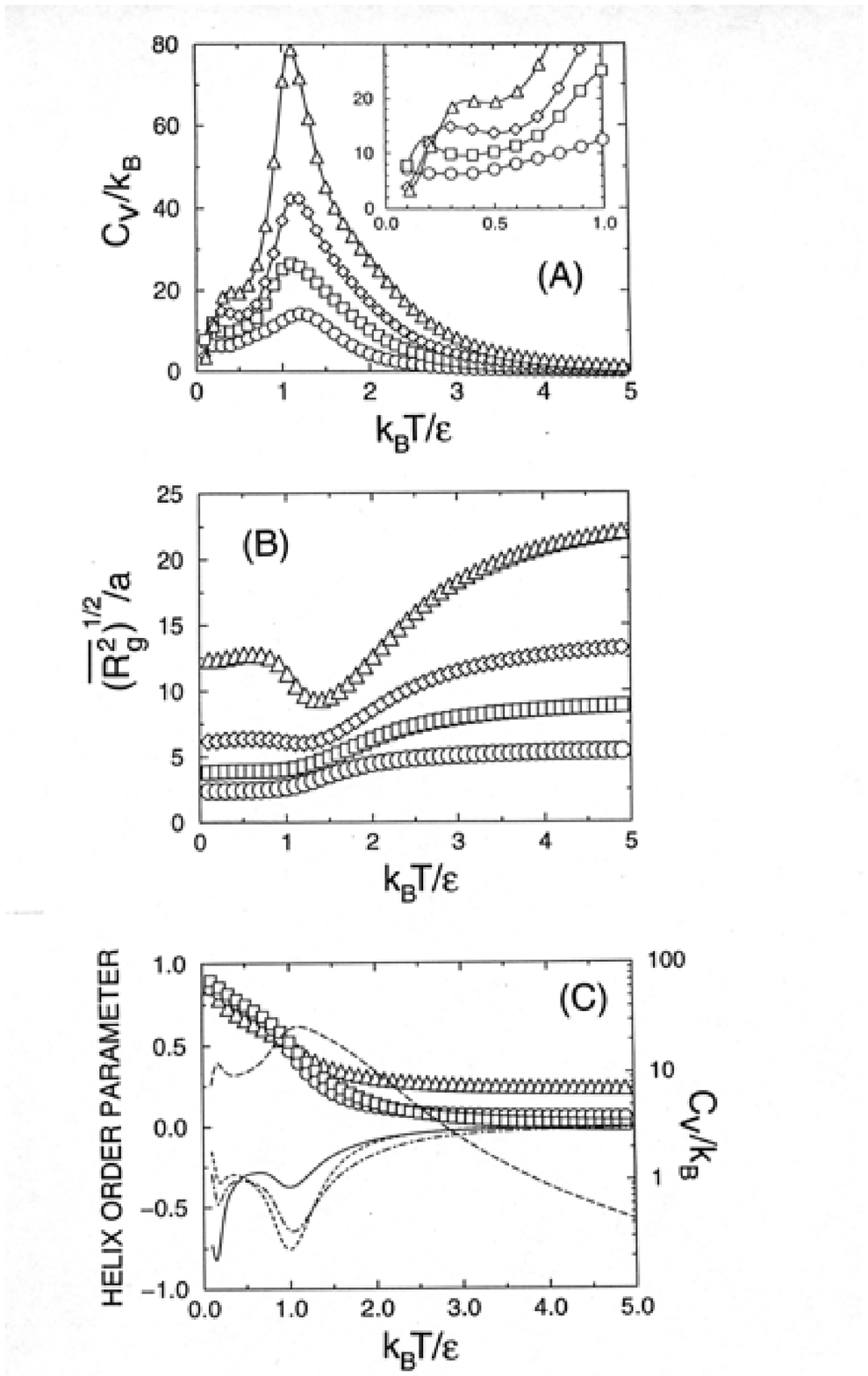}
\caption{Variation of physical quantities as a function of 
scaled temperature $\tilde{T} = k_BT/\epsilon$. Triangles, diamonds, squares, and circles
in Figs 2(a) and 2(b) corresponds to 39mers, 26mers, 19mers, and 13mers
respectively. Only every tenth data point was plotted, and more 
data points are represented by the more smooth curves. The error
bars of the $C_v$ curve are almost twice the  size of the
plotted symbols, and triple the size of the plotted symbols
in the inset.  Triangles, squares and circles
in Fig 2(c) represents the helix  parameters
$H_1$ , $H_2$ and $H_3$ for  the case
of 19mers, respectively. The solid, short-dashed, and the  dot-dashed
curves represent $dH_1/d \tilde{T}$, $dH_2/d \tilde{T}$, and $dH_3/d \tilde{T}$
respectively. An overlay of the  $C_v$ curve (to the right scale) 
is also shown  by the long-dashed curve for reference.}
\label{pic2}
\end{center}
\end{figure}
Returning to Fig. 2a, we would like to 
point out a third feature of interest, which is the emergence of
a shoulder in the larger polymers near $k_BT/\epsilon=2$, 
not easily detectable from Fig. 2a. 
This shoulder corresponds to a $\theta$
temperature collapsing of the polymer chain from a  gyrated
coil (``gaslike")  to a randomly collapsed
globular state (``liquidlike"). This 
collapsing is a 
direct cause of the 
directionally averaged attractive potential between the monomers,
creating globular states which have little
helical ordering. 
In Fig. 2b, the radius of gyration 
curves shows a significant
decrease near a transition 
temperature $k_BT/\epsilon = 2$, as the signature of 
a collapsing transition in polymers. 
At lower temperatures, as helix states start to form,
there is an
increase in both $R_g$ and $R_{\rm end}$ (not shown). 
Such a turnover is actually a characterization 
of the helix formation, because
in a large polymer, the ordered helix 
structure will be quite extended along the helix axis, forcing 
the monomers to have an average radius of gyration larger than
that of a compact collapsed state. 
For shorter helices, the increase of polymer dimensions at low temperature
does not exist, and the coil-globular and globular-helix I transitions
 merge to a single transition. 
Note that neither dimensional parameters show appreciable
changes at the helix I-helix II transition, 
which is again consistent with  our belief
that the helix I to helix II transition 
merely involves the ordering of the end segments and tightening of the
helix segments, with no significant structural
rearrangement.

To summarize, we have described a 
simple model which can be 
used for examining the coil-helix transitions. 
To furnish a stable 
perfect helix ground state, one employs a strongly
directionalized 
attractive potential.
The coil-helix transition actually consists of three steps
as the temperature is lowered.
At high temperatures the
polymer has a gaslike coil 
conformation which collapses into a liquidlike 
globular conformation with the relative bond directions
 still remain disordered. This state then makes a 
transition 
to a relatively well ordered ``solid"-like
helix I state, with a dramatic increase in bond direction 
correlation. 
Upon further cooling this conformation
crystallizes to the very well ordered helix II state. 
It would be interesting to confirm that 
this transition indeed accompanies a $C_v$ anomaly 
experimentally in short polypeptide chains which are known
to exhibit perfect helical ground states. 
We hope the current model will shed some light on 
analyzing experimental data in complimentary to a traditional 
Zimm-Bragg type physical picture.
We believe that
the current treatment can be modified with
little effort to 
produce a variety of ground states,
important to modeling other interesting transitions in biological 
systems.

It is a pleasure to acknowledge insightful conversations with
B. Nickel, D. Sullivan and P. Waldron, the help of C.Z. Cai on plotting 
Fig. 1, 
as well as the financial support of this work by 
the Natural Sciences and Engineering Research Council of Canada. We would also
like to thank M. Gingras for generous CPU time allocation on his Sun Enterprise 450 workstation.


%
%

%
%

\begin{references}
\bibitem{Biobook} See, for example, D. Poland, H.A, Scheraga, {\it Theory of
Helix-Coil Transitions}, Academic Press, New York, 1970; J.M. 
Scholtz and R.L. Baldwin,  {\it Annu. Rev. Biophys. Biomol. Struct.}
{\bf 21}, 95 (1992)


\bibitem{ZB} 
B.H. Zimm, J.K., Bragg, {\it J. Chem. Phys.}, {\bf 30}, 271 (1959).


\bibitem{Theories} See, for example, 
K. Nagai, {\it J. Phys. Soc. Jpn}, {\bf 15}, 407, (1961);
S. Lifson, A. Roig, {\it
J. Chem. Phys.}, {\bf 34}, 1963 (1961);
V. Madison, J. Schellman, {\it Biomolecules}, {\bf 11}, 1041 (1972);
Also  see reference cited in Ref. [1].

\bibitem{Qian} H. Qian and J. A. Schellman, {\it J. Phys. Chem.}
{\bf 96}, 3987 (1992); H. Qian, {\it Biophysical Journal} {\bf 67}, 349 (1994).


\bibitem {Sims} S.-S. Sung, {\it Biophys. Journal} {\bf 68},
826 (1995); D.R. Ripoll, and H.A. Scheraga, {\it Biopolymers} {\bf
27}, 1283 (1988);
S.R. Wilson and W. Cui, {\it Biopolymers}, {\bf 29}, 225 (1990); H. Kawai,
Y. Okamoto, M. Fukugta, T. Nakazawa, and T. Kikuchi,
{\it Chem. Lett.} {\bf 2}, 1991; Y. Okamoto, {\it Proteins Struct. Funct. Genet.} {\bf 19},
14 (1994).

\bibitem{Oka} Y. Okamoto, U.H.E Hansmann, {\it  J. Chem. Phys.} {\bf 99}, 11276 (1995)



\bibitem {Liv} T.B. Liverpool, R.Golestanian, K. Kremer, Phys. Rev. Lett. {\bf 80} (2), 405
(1998)

\bibitem {Zhou}
Y. Zhou, C.H. Hall, M. Karplus, Phys. Rev. Lett. {\bf 77} (13), 2822 (1996)


\bibitem{Muthu} M. Muthukumar, J. Chem. Phys. {\bf 104} (2), 691 (1995)


\bibitem{Yamakawa} H. Yamakawa, {\it Modern Theory of Polymer Solutions}, Harper and Row, New Yrok, (1971).

\bibitem{Pivot} N. Madras, A.D. Sokal, {\it  J. Stat. Phys.} {\bf 50}, 109 (1988) 

\bibitem{Uli} U.H.E Hansmann, Y. Okamoto, {\it Physica A} {\bf 212}, 415 (1994) 

\bibitem{Note} Upon re-examination of the transition curves, obtained by the real potential
model of Okamoto and Hansmann [6], an emergence of a second peak is seen in their data, for 
${\rm Ala_{\rm (20)}}$ which is known to be a strong helix former. However, there is no
discussion given for its existence in Ref. \cite{Oka}.
\end{references}
\end{document}